\documentclass[a4paper,twoside]{article}
\newcommand{\affil}[1]{$^{\rm #1}$}
\usepackage[authoryear]{natbib}
\bibpunct{(}{)}{;}{a}{}{,}
\usepackage{graphicx}
\date{} 
\newcommand{\kms}{\mbox{km\,s$^{-1}$}}
\title{\large\bf\flushleft High Resolution Spectroscopy of the Planetary Nebulae PM\,1-242, PM\,1-318 and PM\,1-322}
\author{\parbox{\textwidth}{\flushleft
\vspace{-0.5cm}
{\it L.F. Miranda\affil{A,B,F}, R. V\'azquez\affil{C}, M.A. Guerrero\affil{A}, C.B. Pereira\affil{D}, and 
E. I\~niguez-Gar\'{\i}n\affil{C,E}   }\\
\vspace{0.4cm}
{\small \affil{A}\,Instituto de Astrof\'{\i}sica de Andaluc\'{\i}a -- CSIC,
  Granada (Spain)}\\
{\small \affil{B}\,Facultade de Ciencias, Universidade de Vigo, Vigo (Spain)
  [Current address] } \\
{\small \affil{C}\,Instituto de Astronom\'{\i}a, Universidad Nacional Aut\'onoma de M\'exico, 
Ensenada (Mexico)}\\
{\small \affil{D}\,Observatorio Nacional -- MCT, Rio de Janeiro (Brazil)}\\
{\small \affil{E}\,Facultad de Ciencias, Universidad Aut\'onoma de Baja California, Ensenada 
(Mexico)}\\
{\small \affil{F}\,Email: lfm@iaa.es}}}
\begin{document}
\twocolumn[
\begin{changemargin}{.8cm}{.5cm}
\begin{minipage}{.9\textwidth}
\vspace{-1cm}
\maketitle
%
%
\small{\bf Abstract: We have recently confirmed the planetary nebula (PN)
  nature of PM\,1-242, PM\,1-318 and PM\,1-322. Here we present high-resolution
  long-slit spectra of these three PNe in order to analyze their
  internal kinematics and to investigate their physical structure. PM\,1-242
  is a tilted ring and not an elliptical PN as suggested by direct images. The object is probably 
  related to ring-like PNe and shows an
  unusual point-symmetric brightness distribution in the ring. PM\,1-318 is a pole-on elliptical PN,
  instead of a circular one as suggested by direct images. PM\,1-322
  is spatially unresolved and its spectrum shows large differences between the forbidden
  lines and H$\alpha$ profiles, with the latter showing a double-peaked profile and relatively 
  extended wings (FWZI $\sim$ 325 km\,s$^{-1}$). These properties are found in other PNe that are suspected 
  to host a symbiotic central star.}


\medskip

\medskip
\end{minipage}
\end{changemargin}
]
\small

\section{Introduction}

Planetary nebulae (PNe) evolve from  Asymptotic Giant Branch (AGB) stars after 
a short post-AGB transition and represent the last evolutionary stage of 
solar-type stars (M $\leq$ 8\,M$_{\odot}$) before the white dwarf phase 
\citep{BLO95}. The transformation of an AGB star into a PN involves a dramatic 
change in all stellar and circumstellar properties \citep{BYF02}. In particular, while 
mass ejection in the AGB phase is spherical, the PN phase is dominated 
by axisymmetric shells, peculiar envelope geometries and collimated outflows 
\citep{MAN96,MRG10}. Direct images allow us to study the morphology of PNe and 
to identify the structural components present in these objects. However, the study of 
the internal kinematics of PNe  is crucial to understand the nature of the components 
observed in the images, to determine their physical structure, and to investigate the 
ejection processes involved in their formation \citep[e.g.,][]{MGT99}.

We have started an intermediate-resolution spectroscopic program to establish 
the true nature of sources classified as post-AGB star and PN candidates on the basis of 
their IRAS colours. Among other results \citep[see][]{PYM07}, we have 
confirmed the PN nature of several objects and analyzed their morphology
\citep[hereafter MPG09]{PYM05,MPG09}. 

In this paper we present high-resolution long-slit spectra of
three of these confirmed PNe: PM\,1-242, PM\,1-318 and PM\,1-322. The spectra
allow us to analyze their internal kinematics and to investigate their
physical structure. At least in two objects, the real
structure derived from the long-slit spectra differs from that suggested by the 
morphology observed in direct images. 

\begin{figure}[h]
\begin{center}
\includegraphics[scale=1.00, angle=0]{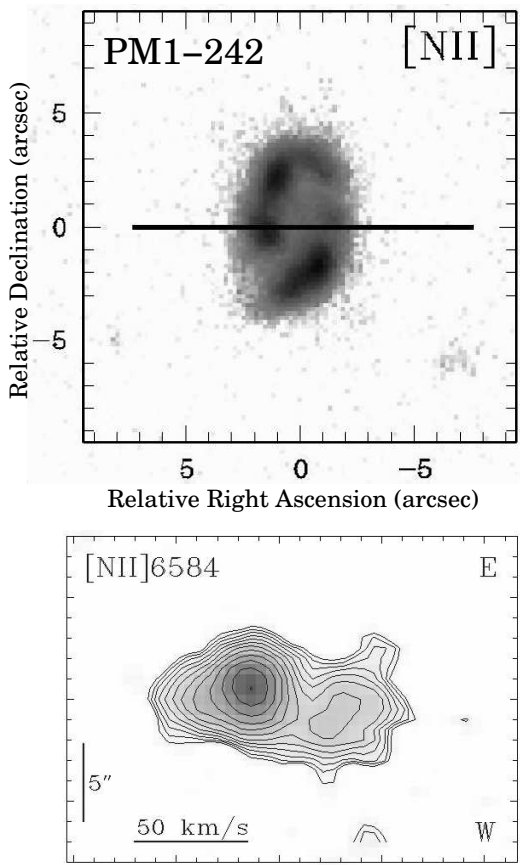}
\caption{({\it top)} [N\,{\sc ii}]$\lambda$6584 image of PM\,1-242 (adapted from MPG09). The slit
  position used for high resolution spectroscopy is indicated. {\it (bottom)}
  Grey-scale and contours position-velocity map of the [N\,{\sc ii}] emission line derived from the
  long-slit spectrum. The orientation (E-W) and the spatial and velocity
  scales are indicated.}\label{fig1}
\end{center}
\end{figure}

\section{Observations}

High-resolution long-slit spectra were obtained using the Manchester Echelle
Spectrometer (MES) at the 2.1\,m telescope on San Pedro M\'artir Observatory
(Mexico) on 2007,
July 16--20. A Site CCD with 1024$\times$1024 pixels was used as detector. In
the case of PM\,1-242 a spectrum in the H$\alpha$ and [N\,{\sc ii}]$\lambda$6584
(hereafter [N\,{\sc ii}]) lines was obtained with the slit oriented at
position angle (PA) 90$^{\circ}$, an exposure time of 900\,s and a 2$\times$2
binning on the detector. For PM\,1-318, the [O\,{\sc iii}]$\lambda$5007
(hereafter [O\,{\sc iii}]) line was observed with the slit oriented at PA
315$^{\circ}$, an exposure time of 1800\,s and without binning. Finally,
in the case of PM\,1-322, the H$\alpha$, [N\,{\sc ii}] and  [O\,{\sc iii}]
lines were observed with the slit oriented at PA 120$^{\circ}$, an exposure
time of 1800\,s and without binning. A slit width of 1.6$''$ was used in all
cases. The long-slit spectra were flat fielded, bias corrected and, finally,
wavelength calibrated with a Th-Ar lamp to an accuracy of $\pm$1
km\,s$^{-1}$, using standard routines for long-slit spectroscopy within the IRAF package. The
spectral resolution, determined by the FWHM of the Th-Ar comparison lines, is
$\simeq$ 12\,km\,s$^{-1}$. The spatial resolution, determined by the
seeing, is 1$''$--1.5$''$. 

\section{Results and Discussion}

In the following, we will present and discuss the results obtained for the
three PNe. A brief description of their morphology
and spectral properties is also provided from the previous
results by \cite{PYM05} and MPG09.

\subsection{PM\,1-242 (IRAS\,18320+0005)}

PM\,1-242 is a medium-to-high excitation PN with relatively strong He\,{\sc
  ii} emission and relatively weak [N\,{\sc ii}] and [S\,{\sc ii}] 
emissions (see MPG09). Figure\,1 shows the [N\,{\sc ii}] image of PM\,1-242
presented by MPG09. Morphologically, PM\,1-242 can be classified as 
an elliptical PN with its major axis along the north-south direction and a
size of $\simeq$ 8$''$$\times$5$''$. 
Two bright low-excitation, point-symmetric arcs trace the ellipse. Faint
protrusions and/or knots are observed along PAs 
50$^{\circ}$, 225$^{\circ}$ and  230$^{\circ}$, particularly in H$\alpha$ and
[O\,{\sc iii}] (see MPG09). As mentioned by MPG09, the
images do not allow us to determine whether PM\,1-242 is an ellipsoid
containing two point-symmetric arcs, with the major 
axis along the north-south direction, or a bipolar PN with a bright ring, faint 
bipolar extensions represented by the protrusions/knots and with the major
axis along the east-west direction.  

A position-velocity (PV) map of the [N\,{\sc ii}] line derived from the
long-slit spectrum is also shown in Figure\,1. The PV map of the H$\alpha$
line (not shown here) 
is qualitatively similar to the [N\,{\sc ii}] one. The PV maps show two
velocity components separated $\simeq$ 36 and 
$\simeq$ 42 km\,s$^{-1}$ in H$\alpha$ and [N\,{\sc ii}], respectively. The two
components correspond to emission from the 
arcs at the minor axis of the ellipse (Figure\,1). In particular, the
blueshifted bright component corresponds to a bright knot at the eastern egde
of the ellipse while the redshifed faint component corresponds to a faint knot
at the western edge. From the center of the H$\alpha$ and [N\,{\sc ii}] emission 
features in the PV maps we derive a heliocentric systemic velocity of $\simeq$ 
$+$72.6 {\kms}. 

The presence of two velocity components at different velocities is not
compatible with the slit being oriented along the minor axis of an ellipsoid. If this were 
the case, one would expect the two components to be located at the systemic
velocity. On the contrary, the results strongly suggest that the observed ellipse is a
tilted ring with its polar axis oriented along the east-west direction. The ring is
tilted so that its eastern side points toward the observer and
its western side recedes away. Assuming that the ring is circular, we derive an
inclination angle of $\simeq$ 30$^{\circ}$ for its polar axis with respect to
the observer. Taking into account this angle, the
expansion velocity of the ring is $\simeq$ 20 and $\simeq$ 25 km\,s$^{-1}$ in 
H$\alpha$ and [N\,{\sc ii}], respectively. 

These results show that the main structure of PM\,1-242 is a tilted ring. The
faint protrusions and knots (see MPG09) suggest that faint bipolar
lobes may also be present and, hence, PM\,1-242 could be a bipolar PN,
although the presence of faint lobes needs to be confirmed by deeper
observations. In any case, PM\,1-242 is probably related to ring-like PNe like, e.g.,
IC\,2149 \citep{VAZ02}, Me\,1-1 \citep{PER08}, WeBo\,1 \citep{BPW03} and
SuWt\,2 \citep{JON10}. In these PNe, a bright ring is the dominant
structure in the nebula, which is, in some cases, accompanied by very faint
bipolar lobes. Particularly interesting in PM\,1-242 is the point-symmetric
brightness distribution in the ring. Point-symmetry is usually observed in the
lobes of bipolar PNe or in collimated structures usually associated to these
objects, but to the best of our knowledge PM\,1-242 is the only PN that shows
this characteristic in the equatorial structure.

\begin{figure}[h]
\begin{center}
\includegraphics[scale=0.78, angle=0]{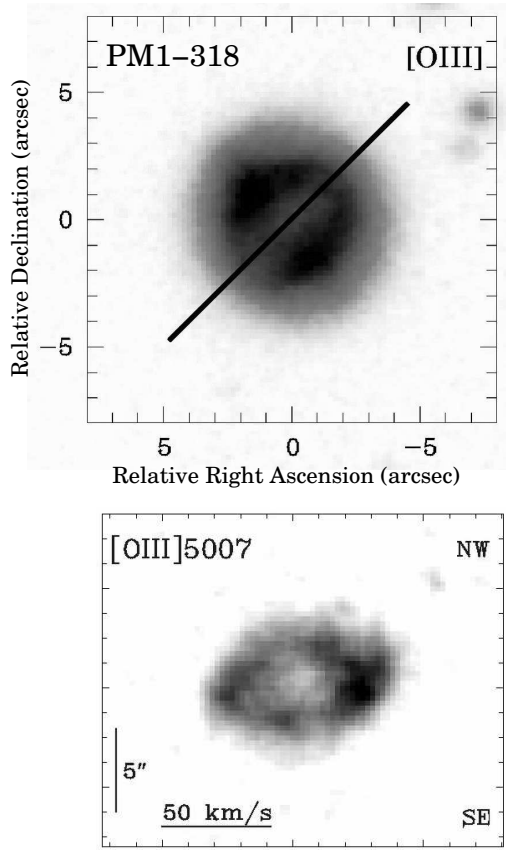}
\caption{{\it (top)} [O\,{\sc iii}]$\lambda$5007 image of PM\,1-318 (adapted 
from MPG09). The slit position used for high 
resolution spectroscopy is indicated. {\it (bottom)} Grey-scale
position-velocity map of the [O\,{\sc iii}] emission line derived 
from the long-slit spectrum. The orientation (NW-SE) and the spatial and
velocity scales are indicated.}\label{fig2}
\end{center}
\end{figure}

\begin{figure}[h]
\begin{center}
\includegraphics[scale=0.7, angle=0]{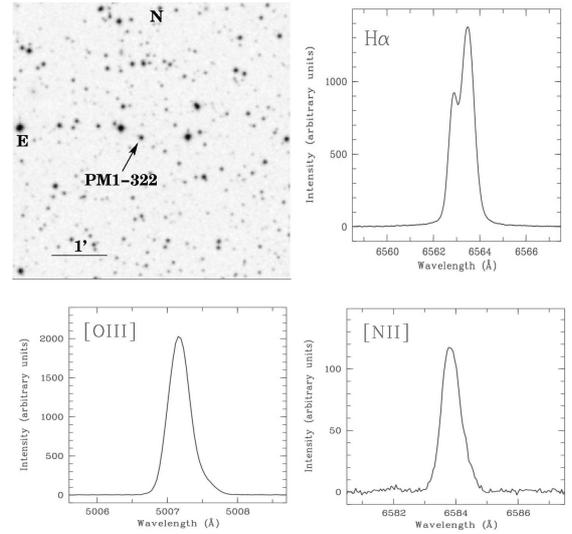}
\caption{Line profiles of the H$\alpha$, [N\,{\sc ii}] and [O\,{\sc iii}]
  emission lines in PM\,1-322. An identification chart obtained from the POSS
  is also shown (upper left).}\label{fig4}
\end{center}
\end{figure}

\subsection{PM\,1-318 (IRAS\,20077+3722)}

PM\,1-318 is a high-excitation PN with strong He\,{\sc ii} lines
(e.g., He\,{\sc ii}$\lambda$4686/H$\beta$ $\simeq$ 1) and extremely weak or
absent low-excitation emission lines (MPG09). The [O\,{\sc iii}] image of PM\,1-318
by MPG09, reproduced in Figure\,2, shows that the nebula consists of an inner ring-like
shell with two enhanced opposite regions, surrounded by a faint, round   
attached shell. The morphology observed in the images classifies PM\,1-318 as
a round PN.

A PV map of the [O\,{\sc iii}] emission line is 
also presented in Figure\,2. The [O\,{\sc iii}] emission feature corresponds
to emission from the inner shell because the spectrum is not deep enough to detect
the faint outer shell. The [O\,{\sc iii}] emission feature appears as a tilted velocity
ellipse on the PV map. The southeastern nebular regions are blueshifted while
the northwestern nebular regions are redshifted. A velocity splitting of $\simeq$ 60 
km\,s$^{-1}$ is measured at the center of the emission feature and a heliocentric systemic 
velocity of $\simeq$ $-$99 {\kms} is derived from that center. We note that radial velocities up to $\simeq$
$\pm$ 40 km\,s$^{-1}$ from the systemic velocity are observed. The inner regions of the
velocity ellipse show a decreasing [O\,{\sc iii}] intensity, which suggests
that the [O\,{\sc iii}] emission mainly originates in a relatively thin layer
of the inner shell, as also suggested by the direct image. 

The tilt of the velocity ellipse on the PV map rules out that the inner shell
of PM\,1-318 is spherical (or round) because, in this case, the velocity
ellipse would not be tilted on the PV map. This result strongly suggests that the
inner shell is an ellipsoid with the major axis tilted with respect to the
observer and oriented near PA 315$^{\circ}$. Moreover, the largest velocities
in the velocity ellipse are observed almost along the line of sight to
the center of the nebula. This implies a relatively small inclination angle of
the polar axis with respect to the
observer and, hence, a polar expansion velocity of the order of 
$\sim$ 30--40 km\,s$^{-1}$. From these results we conclude that PM\,1-318 is
an (almost) pole-on elliptical PN. The enhanced regions in the inner shell
probably represent a bright equatorial region in the ellipsoid.

\subsection{PM\,1-322 (IRAS\,20124+5844)}

PM\,1-322 was identified as a new PN by \citet{PYM05}. It exhibits very strong
[O\,{\sc iii}]$\lambda$4363 line emission ([O\,{\sc iii}]$\lambda$4363/H$\beta$ $\simeq$ 1.2) 
indicating a high electron density ($\geq$ 10$^6$ cm$^{-3}$). The nebula is unresolved in our direct
images. The compactness and high electron density suggest that
PM\,1-322 is a very young PN. On the other hand, these properties and the 
position of PM\,1-322 in the [O\,{\sc  iii}]$\lambda$4363/H$\gamma$ versus [O\,{\sc
  iii}]$\lambda$5007/H$\beta$ diagram indicate that PM\,1-322 could be a 
symbiotic star \citep{PYM05}.

PM\,1-322 is also spatially unresolved in the long-slit spectra. Therefore, we
show in Figure\,3 the H$\alpha$, [N\,{\sc ii}] and [O\,{\sc iii}] emission
line profiles derived from the spectra. The  [N\,{\sc ii}] and [O\,{\sc iii}] profiles 
presents a single peak at a heliocentric radial velocity of $\simeq$ $+$27.2\,{\kms} that 
can be considered as the systemic velocity of the nebula. The FWHM (corrected of instrumental 
resolution) of the [N\,{\sc ii}] and [O\,{\sc iii}] profiles 
is $\simeq$ 30 and $\simeq$ 18 km\,s$^{-1}$, respectively. The H$\alpha$ emission line shows a 
double-peaked profile. The heliocentric radial velocity derived from the centroid of the H$\alpha$ 
profile is $\simeq$ $+$27.2\,{\kms}, in excelent agreement with the systemic velocity of PM\,1-322 
derived from the forbidden lines. The two emission peaks of the H$\alpha$ profile 
are observed at $-$17.5 and $+$9\,{\kms} with respect to the systemic velocity and the red peak is 
stronger than the blue one. The emission peaks are separated by an apparent ``absorption reversal'' that 
is blueshifted by $-$9.6\,{\kms} with respect to the systemic velocity. Finally, the wings of 
the H$\alpha$ line can be traced up to $\simeq$ 325\,km\,s$^{-1}$ (FWZI) while 
the [N\,{\sc ii}] and [O\,{\sc iii}] lines present a much smaller FWZI of
$\simeq$ 90\,km\,s$^{-1}$.

The large differences between the forbidden lines and H$\alpha$ profiles suggest a different 
origin for these emissions. Given the high density in
the object, the regions probed via the H$\alpha$ emission may critically depend upon the
density and its variation within the object. Therefore, the H$\alpha$
emission may be produced in regions that do not contribute to [N\,{\sc ii}]
and [O\,{\sc iii}] emissions. In particular, the relatively large
wings of the H$\alpha$ emission could be due to Rayleigh-Raman scattering in a very
dense region close to the central star \citep[see][]{LEE00}. The spectral properties of 
PM\,1-322 are remarkably similar to these found in other PNe suspected to host a symbiotic central star. 
These PNe are characterized by differences between the forbidden lines and H$\alpha$ profiles, a double-peaked 
H$\alpha$ profile, with very similar properties to these observed in PM\,1-322, and extended H$\alpha$ 
wings \citep[e.g.,][]{BAL89,MTE96,GUE01,AYT03}. These results reinforce the idea that the central star 
of PM\,1-322 is a symbiotic star.

\subsection{Conclusions}

We have presented high-resolution long-slit spectra of three new confirmed 
PNe: PM\,1-242, PM\,1-318 and PM\,1-322 with the aim of investigating their
internal kinematics and physical structrure. The main conclusion of this work
can be summarized as follows.

1) PM\,1-242 is a ring-like PN, but not an elliptical PN as suggested by
direct images.Interestingly, the ring displays a point-symmetric brightness 
distribution which is unusual among ring-like PNe and bipolar PNe with central rings.

2) PM\,1-318 is an almost pole-on elliptical PN with an enhanced equatorial
region, but not a round PN as suggested by direct images. 

3) PM\,1-322 is spatially unresolved and presents large differences between 
the forbidden lines and H$\alpha$ profiles, a double-peaked H$\alpha$ profile and relatively 
extended H$\alpha$ wings. These characteriscs are also found in other PNe believed to
host symbiotic central stars.

\section*{Acknowledgments}

This work has been supported partially by grants AYA2005-01495 of the Spanish
MEC, and AYA2008-01934 of the Spanish MICINN, by grant FQM1747 of
Consejer\'{\i}a de Innovaci\'on, Ciencia y Empresa of Junta de
Andaluc\'{\i}a, and by CONACYT grant 102582 and PAPIIT-UNAM 
grant IN109509 (Mexico).


\end{document}